\documentclass[12pt]{iopart}
\usepackage{caption}
\usepackage{subcaption}
\usepackage{graphicx}

\begin{document}

\title[Cellular Automata model for period-$n$ synchronization]{Cellular Automata model for period-$n$ synchronization: A new universality class.}

\author[1]{Divya D. Joshi \& Prashant M. Gade}
\ead{prashant.m.gade@gmail.com}
\address{Department of Physics, Rashtrasant Tukadoji Maharaj Nagpur University, Amravati Road, Ram Nagar, Maharashtra, Nagpur, India-440033}
\vspace{10pt}

\begin{abstract}
There are few known universality classes of absorbing phase transitions 
in one dimension and most models fall in the well-known directed 
percolation  (DP)  class. Synchronization is a transition to an 
absorbing state and this transition is often DP class. With local 
coupling, the transition is often to a fixed point state. Transitions 
to a periodic synchronized state are possible.
We model those using a cellular automata model with states 1 to $n$. 
The rules are a) Each site in state $i$ changes to state $i+1$ for 
$i<n$ and 1 if $i=n$. 
b) After this update, it takes the value of either neighbour unless 
it is in state 1. With these rules, we observe a transition 
to synchronization with critical exponents different from those of 
DP for $n>2$. For $n=2$, a different exponent is observed.
\end{abstract}

%
%
%
%
%

\section{Introduction}
The theory of phase transitions is a major success story in statistical 
physics in the 70’s. Though the concept of universality was known, the concept of renormalization group was introduced in those times. It was fascinating 
that the second-order phase transition in quark-gluon plasma is in the 
same universality class as the Ising model \cite{rischke2004quark}. Thus, 
seemingly different systems arising in very different contexts are in 
the same universality class depending mainly on the dimension and symmetries \cite{stanley}. 
There has been huge progress in phase transitions in equilibrium 
systems. This led to studies in several nonequilibrium systems 
from the phase transition viewpoint. Among the nonequilibrium transitions, 
transitions to absorbing states are most extensively investigated \cite{hinrichsen2000non}.
These transitions are in a few different universality classes 
such as directed percolation (DP), directed Ising, dynamical
percolation, Manna class etc. \cite{henkel2011non}. Whether or not the 
pair contact process with diffusion (PCPD) is a new universality class 
has been a long-standing problem with no clear solution in 
sight \cite{henkel2004non, gredat2014finite}.
The essential limitation is that the Monte-Carlo simulations are for 
finite-size lattices for finite time and can lead to uncertain conclusions.
The overall evidence in the recent past indicates that PCPD is 
likely to be in the same universality class as a DP in the thermodynamic 
and asymptotic limit \cite{matte2016persistence}.

\begin{figure}[h]
\centering 
\includegraphics[width=0.49\textwidth]{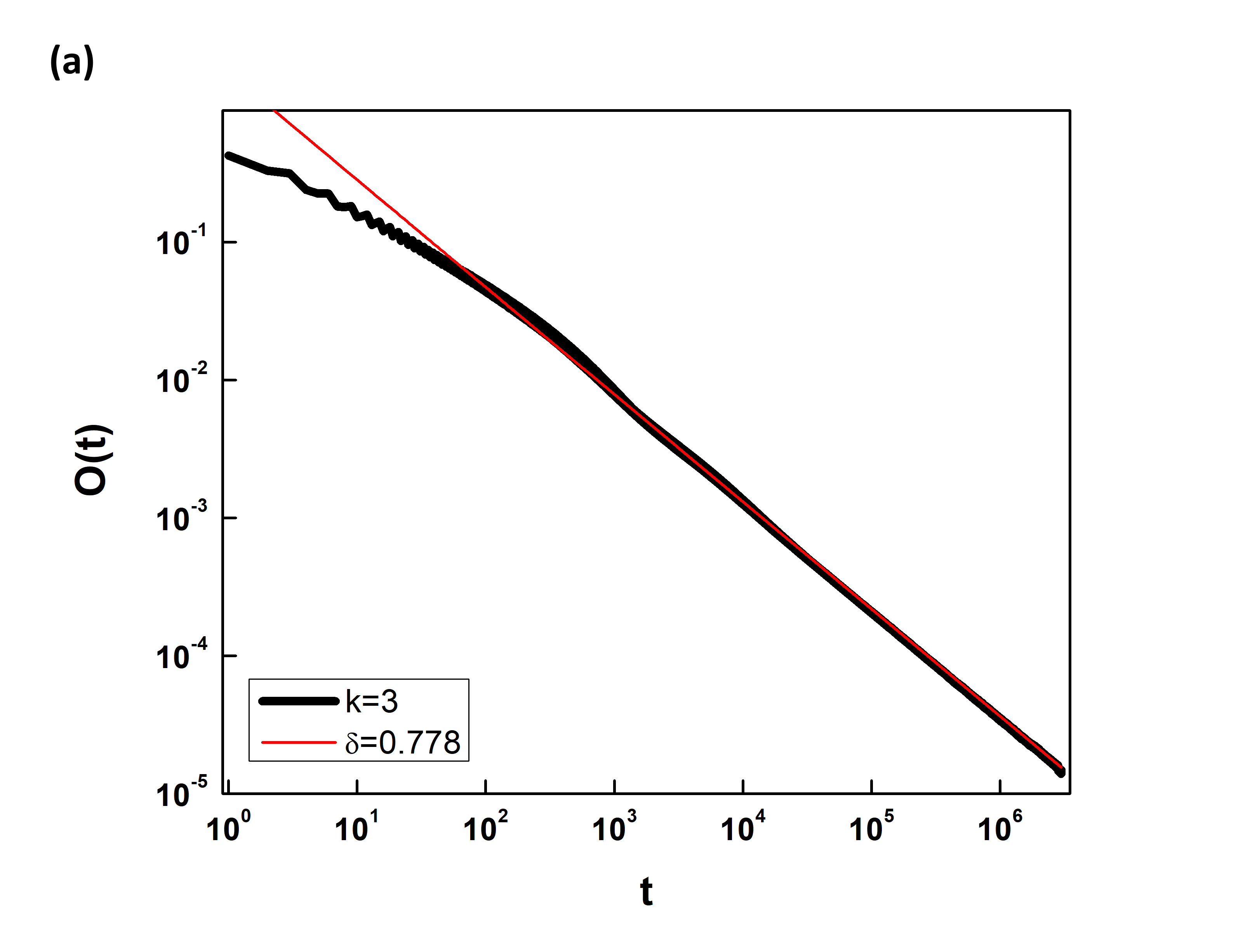} 
\includegraphics[width=0.49\textwidth]{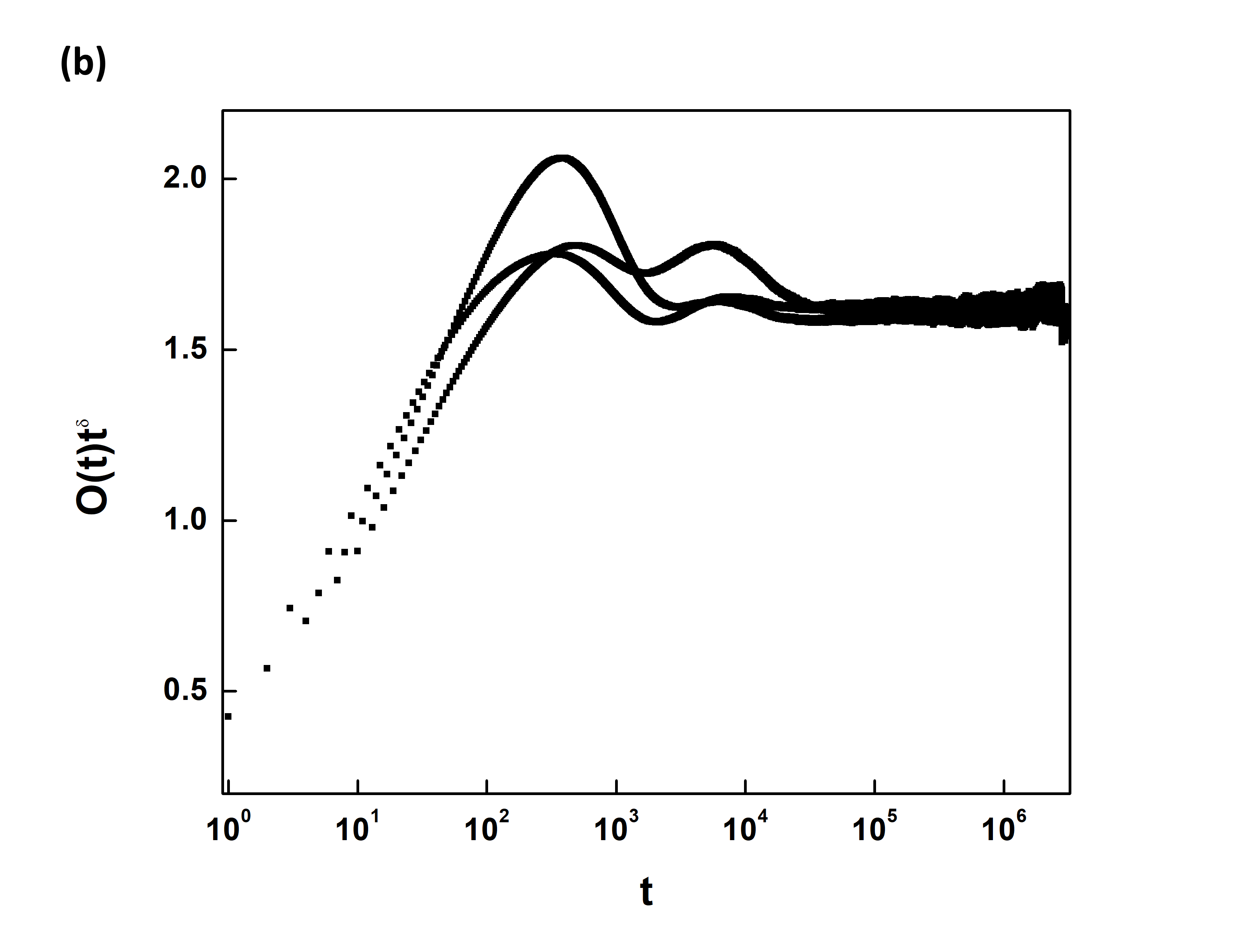} 
\caption{a) Plot for $O(t)$ vs $t$ for states $k=3$, $N=10^6$. We average over 1200 configurations. The decay exponent for this case is $\delta=0.778$. b) We plot $O(t) t^{\delta}$  as a function of $t$. Asymptotically, we obtain a constant value.}
\label{fig1}
\end{figure}
\begin{figure}[h]
\centering 
\includegraphics[scale=0.35]{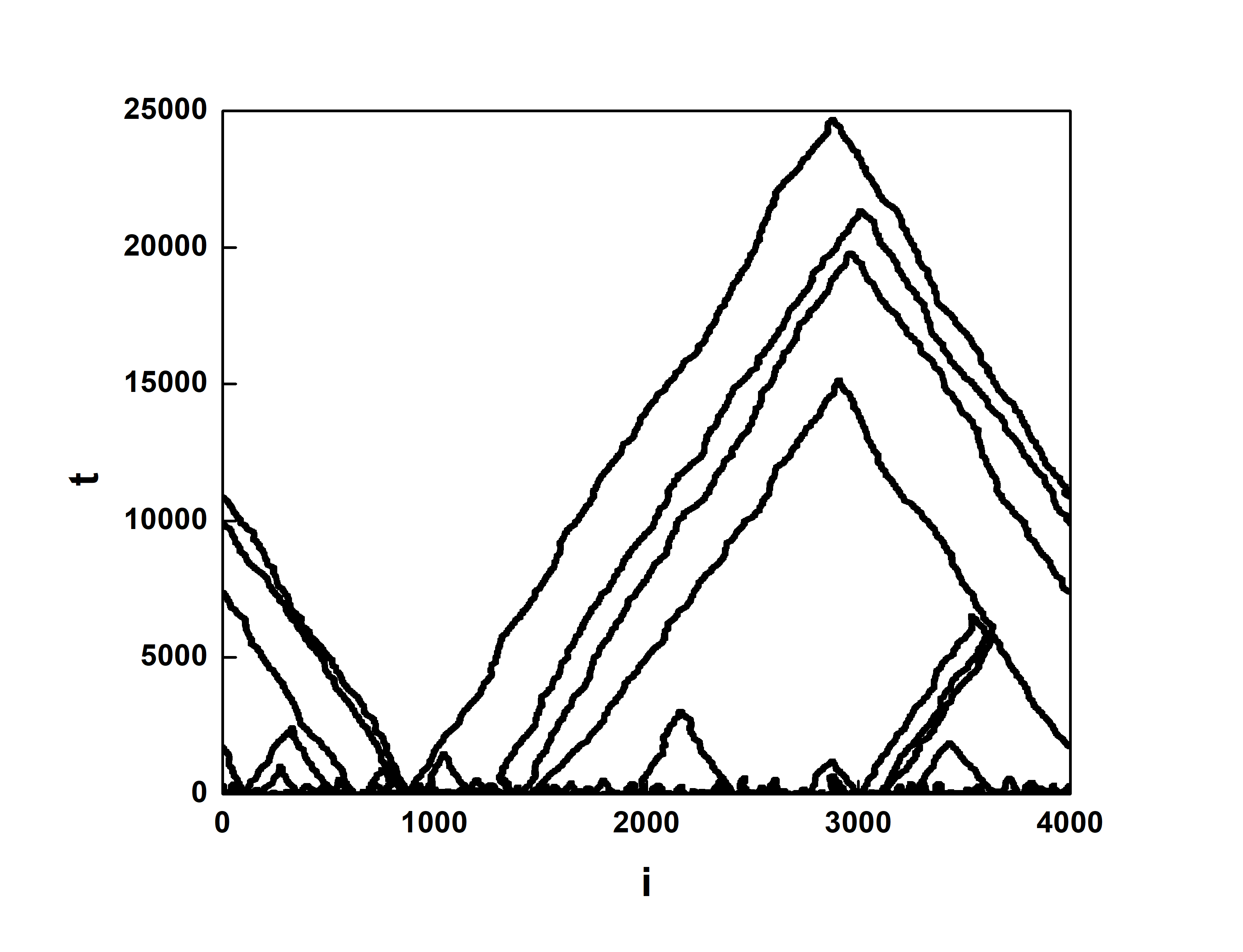} 
\caption{Defect dynamics of the model for $N=4000$, $k=3$. The dynamics for this differ from that of directed percolation.} 
\label{fig2}
\end{figure}
\begin{figure}[h]
\centering 
\includegraphics[width=0.5\textwidth]{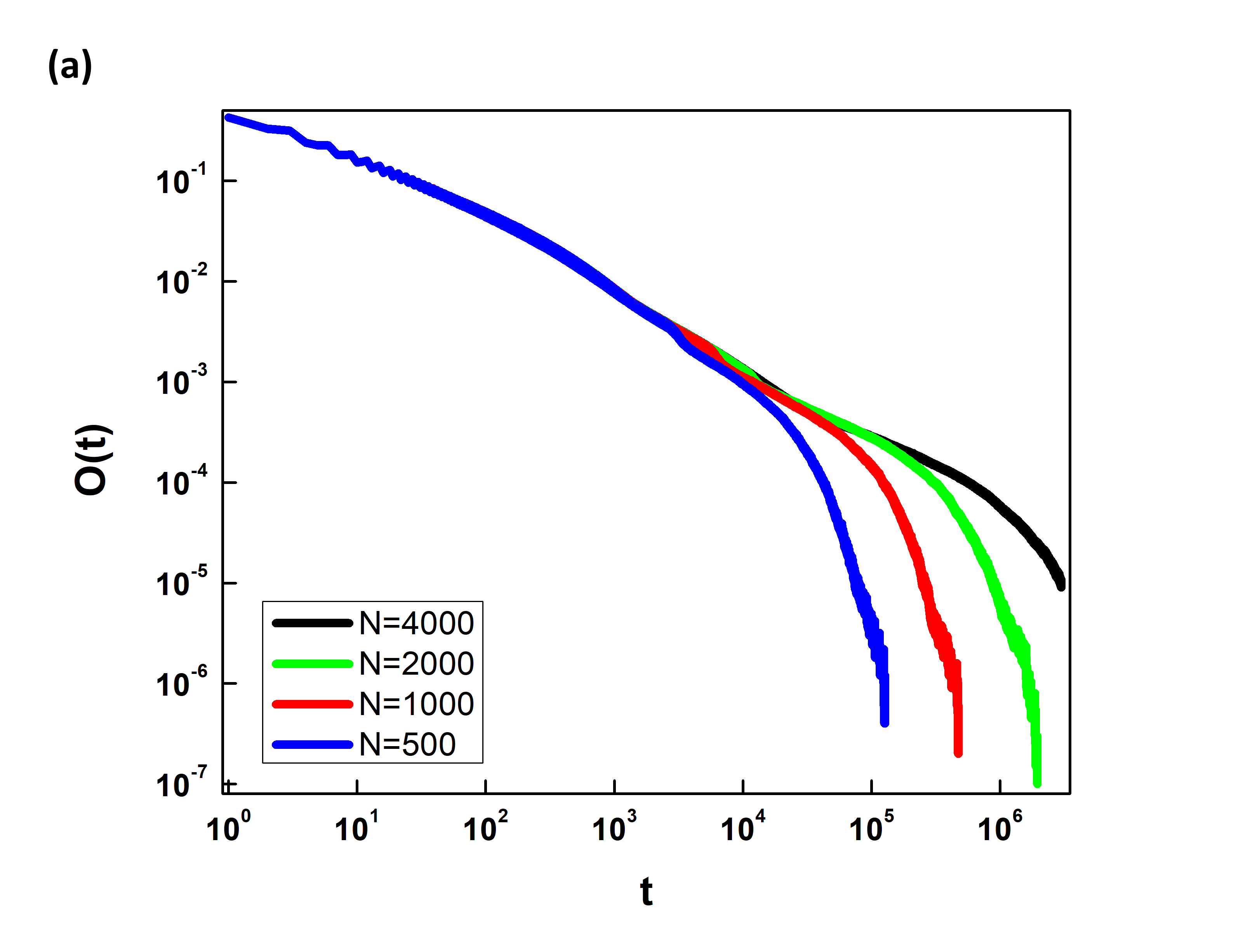} \\
\includegraphics[width=0.49\textwidth]{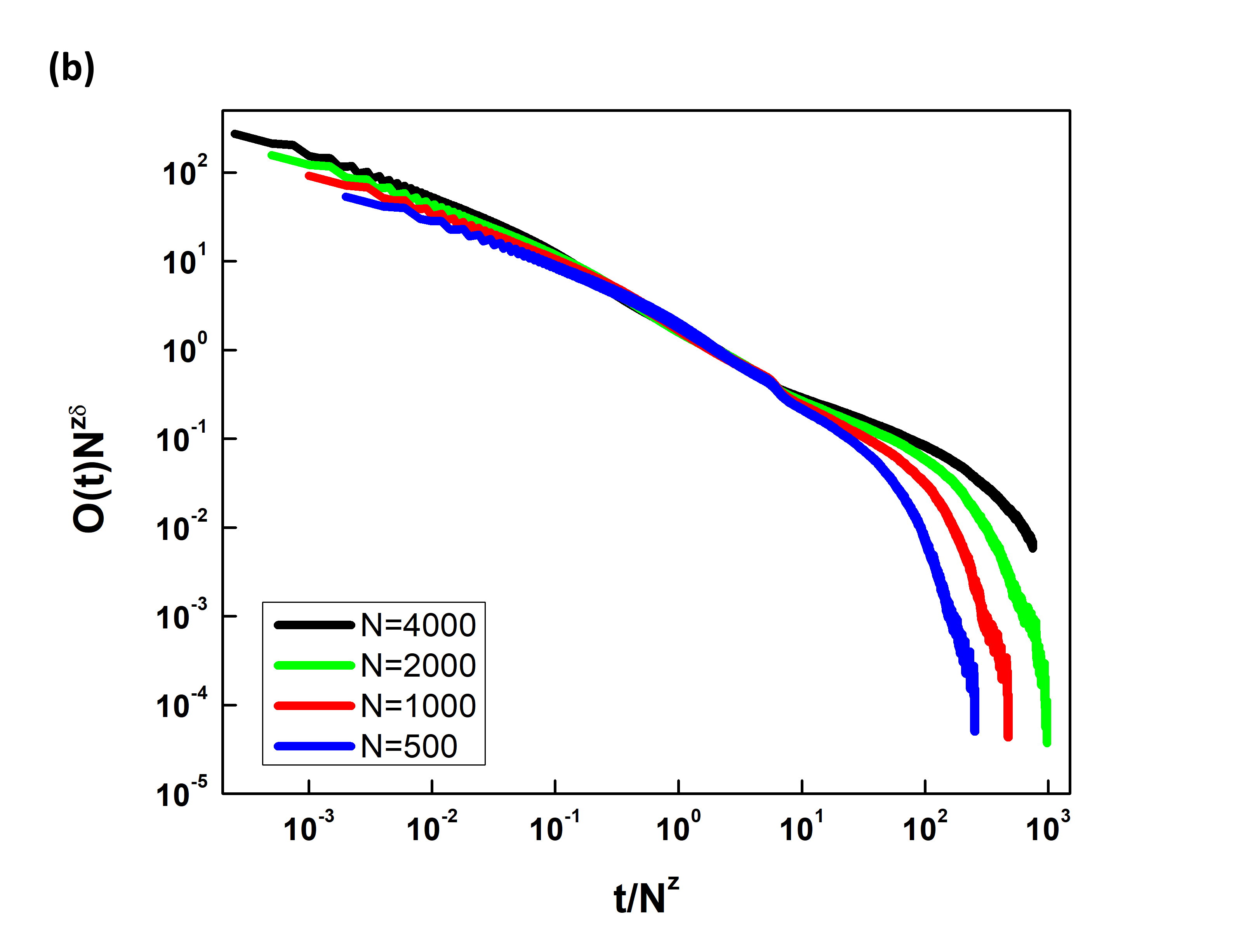} 
\includegraphics[width=0.49\textwidth]{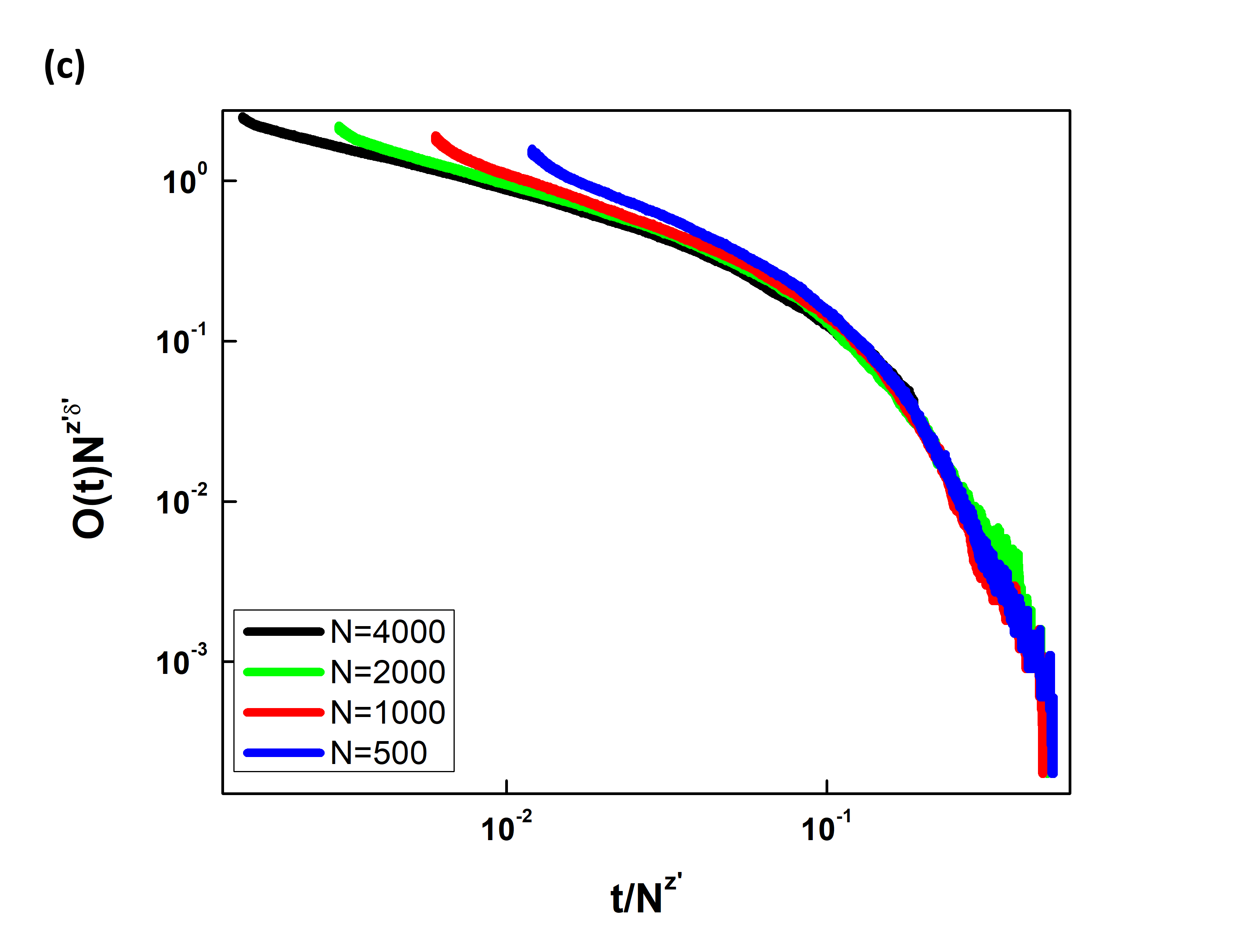} 
\caption{a) We plot $O(t)$ versus $t$ for $N=4000,2000,1000$ and 500 with $k=3$.  We observe a power-law decay with 0.778 followed by a cusp. The cusp is observed approximately for $t=6N$ for all sizes. Above the cusp,  we observe power-law decay with
different power and decay.
b) Plot $O(t)N^{z\delta}$ vs $tN^{z}$ shows good collapse till the cusp.  We observe the excellent collapse 
fo $z=1$ and $\delta=0.778$.
c) Plot $O(t)N^{z'\times \delta'}$ vs $tN^{z'}$ demonstrates a nice collapse for the values $t>6N$, \textit{i.e.}, after the cusp. Here the exponents are $z'=2$ and $\delta'=0.5$.} 
\label{fig3}
\end{figure}

\begin{figure}[h]
\centering 
\includegraphics[scale=0.35]{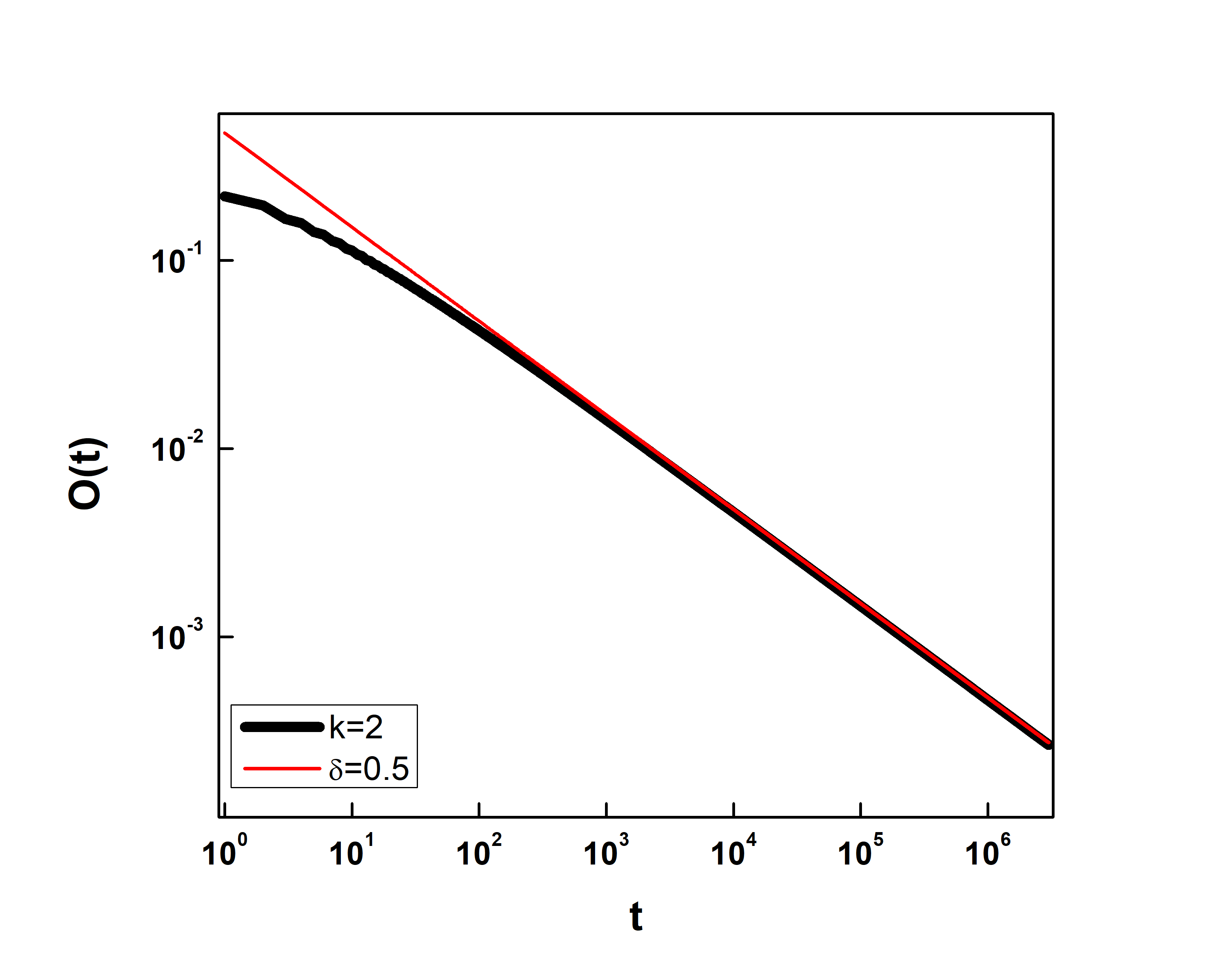} 
\caption{Plot for $O(t)$ vs $t$ for states $k=2$, $N=10^6$ and $t=3\times10^6$. The decay exponent for this case is $\delta=0.5$.} 
\label{fig5}
\end{figure}
\begin{figure}[h]
\centering 
\includegraphics[scale=0.35]{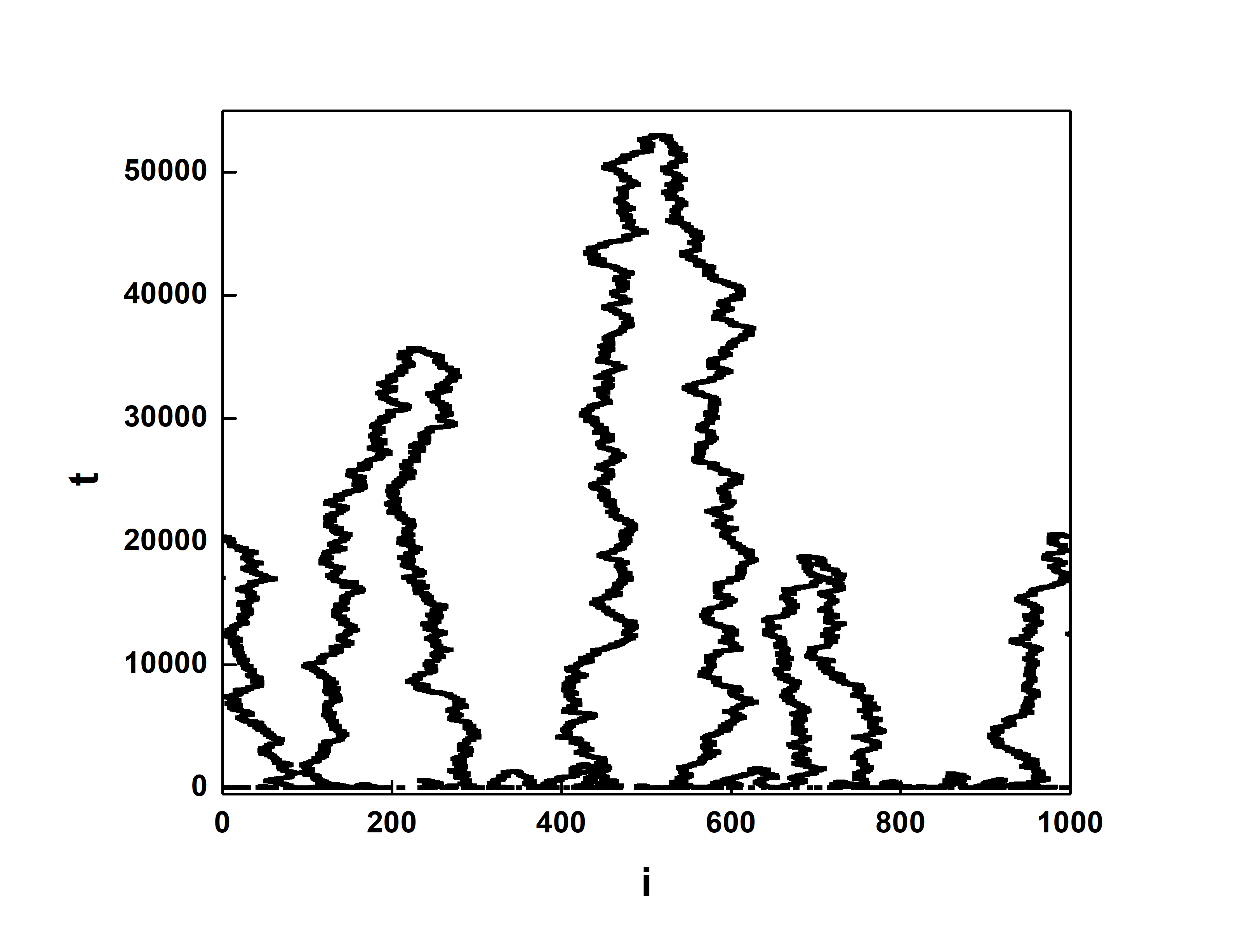} 
\caption{Defect dynamics of the model for $N=1000$, $k=2$. In this case, the dynamics of the model random walk.} 
\label{fig6}
\end{figure}
\begin{figure}[h]
\centering 
\includegraphics[scale=0.3]{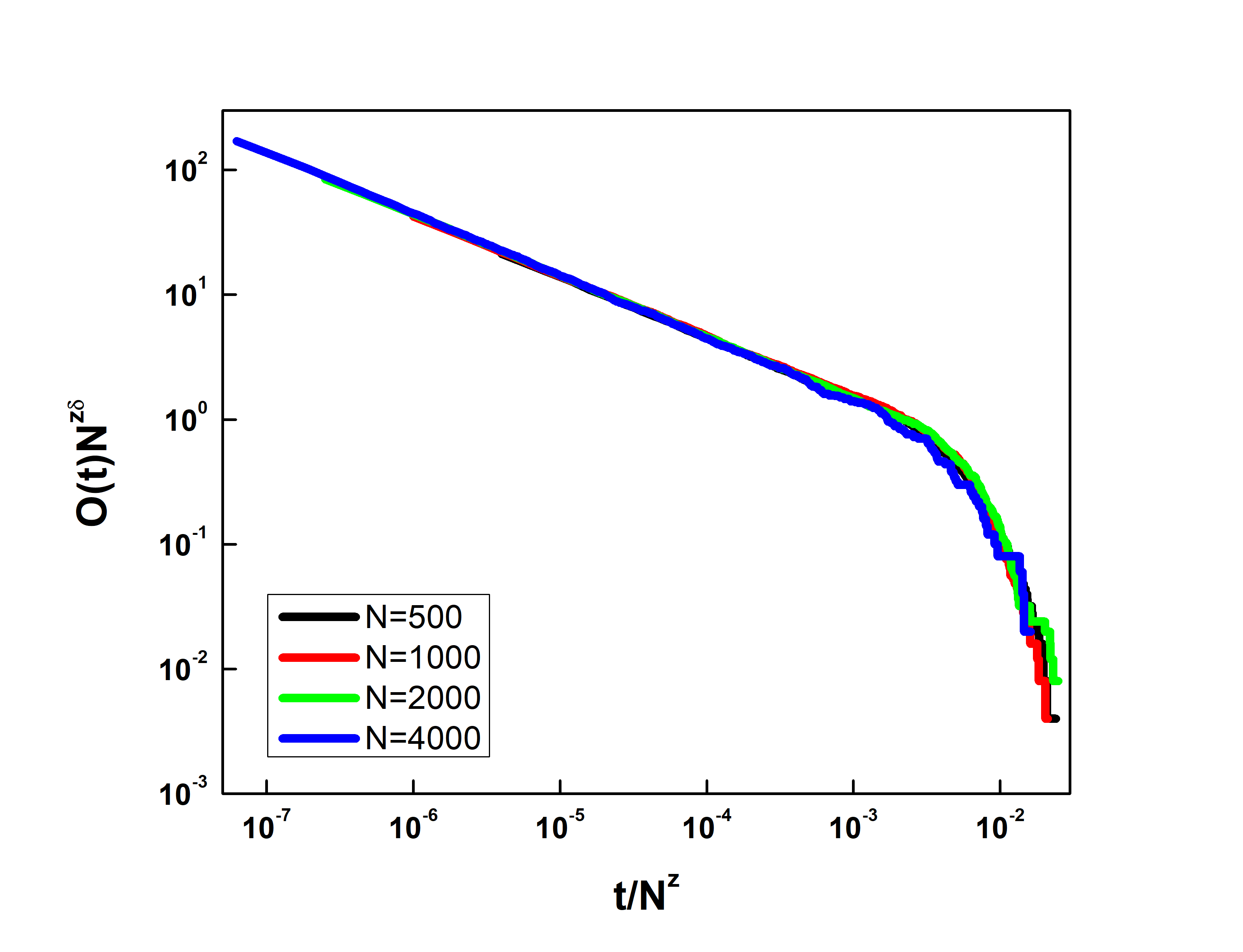} 
\caption{Plot $O(t)N^{z\delta}$ vs $t/N^{z}$ for different size of lattice \textit{i.e.}, $N=500,1000,2000$ and 4000. Here $k=2$, $\delta=0.5$ and $z=2$.} 
\label{fig7}
\end{figure}

Of these different universal classes, the DP 
universality class is the most observed. As the name suggests,
it was initially proposed to model liquid percolating in a medium
under the influence of the gravitational field. Thus there is a preferred direction. 
This preferred direction can be temporal because we cannot go back in time.
Hence this transition can be applicable in dynamic transitions. The majority of these 
transitions are transitions to a unique absorbing state. Now, we are aware of several 
models where DP transition is observed. It is observed 
in contact processes \cite{contact}, Ziff-Gulari-Barshad model for catalysis \cite{zgb}, Domany-Kinzel cellular 
automata \cite{domany} etc. It is proposed to be relevant even for the
transition to turbulence \cite{turbulence}. The Jannsen-Grassberger 
conjecture states that  “the universality class of DP contains 
all continuous transitions from a dead or absorbing state to an 
active one with a single scalar order parameter, provided the dead state is not degenerate and provided some technical points are 
fulfilled: short-range interactions both in space and time, the 
nonvanishing probability for any active state to die locally, 
translational invariance (absence of frozen randomness), and absence of multicritical points” \cite{grassberger1981phase}. 
However, even if the above conditions are relaxed the DP transition is observed \cite{bhoyar2022robustness}.
It is observed even in the most unexpected places where the 
absorbing state has no long-range order in space. 
The first such example is the pair contact process (PCP)
that has infinite absorbing states and still shows the transition 
in DP class \cite{jensen1993critical}. More such examples can be found
in \cite{pakhare2020novel,gaiki2024transition}.
 Another class proposed by Grassberger and coworkers is directed Ising transition 
where the absorbing state is doubly degenerate linked
by symmetry \cite{grassberger1984new}. This universality class was later observed in 
other models such as branching-annihilating random walk \cite{bramson}, interacting
monomer-dimer model \cite{park1,park2,hwang}, nonequilibrium kinetic Ising model \cite{menyhard} and even nonlinear
map with delay \cite{mahajan}. It is a nonequilibrium model and has exponent $\sim$ 0.285 at the critical point \cite{frojdh2001directed}. Another class known as the voter class is 
proposed as a model for opinion formation in society. In the voter universality class,
each agent is initially assigned two opinions. The update rule is
to choose a site randomly and assign it the opinion of
a randomly chosen neighbour \cite{dornic}. In one dimension, it is
equivalent to compact directed percolation \cite{henkel2011non}
and can be mapped to an equilibrium model. 

In nonlinear maps,
along with period-doubling cascade, there can be attractor 
merging critical points. For example, the logistic map
can show a two-band attractor. The bands are two unconnected pieces 
in the phase space that are visited periodically. These bands
gradually increase in size and a band-merging crisis occurs
when it collides with an unstable periodic orbit \cite{ott2002chaos}.
In a coupled map lattice
which is a spatially extended system comprising of such maps,
we can observe a multi-band attractor and a band-merging
crisis. These bands can have long-range order in space. For instance,
consider a state where all even sites are in the same band and all
odd sites are in another band. We can view it as an antiferromagnetic long-range order.
The band merging
transition in this case is in Ising class or directed
Ising class \cite{gade2013universal,gaiki2024transition,mahajan}.
However, when there is no long-range
spatial order, there are infinite possible absorbing states in the thermodynamic limit.
In this case, a band-merging transition is in 
DP class \cite{pakhare2020novel,gaiki2024transition}. This is yet another case of 
DP transition for the highly degenerate absorbing state. 

In our model, the sites tend to take the same value as their neighbours and it can
be described as ferromagnetic coupling. The Ising model is a useful starting point for a system where
individual units can have two states.
Similarly, the Potts model is a useful starting
point for a system where individual units can be in 
 $q$ different states. It shows rich behaviour even in one dimension.
(See Fig. 1 of \cite{dhar1995relaxation}.) Potts model in one dimension leads to $t^{-1/2}$ decay of energy density
for Glauber dynamics and $t^{-1/3}$ decay of energy density for Kawasaki dynamics
for any $q$ for ferromagnetic coupling. (This is equivalent to defect density in our model.) 
For antiferromagnetic coupling.
Potts model for $q=2$ in one dimension leads to $t^{-1/2}$ decay
for Glauber dynamics and $t^{-1/4}$ decay for Kawasaki dynamics.
For higher values of $q$, the energy density
does not decay as a power-law. The voter model
falls in the class of models with short-range interaction and non-conserved order parameters. 
The domains grow as $t^{1/2}$ in this case \cite{derrida}.
 We note that the power-law exponents, if any, 
are different from the one obtained in this work.

We consider a model that can be considered analogous
to the $q$-Potts model with synchronous Glauber dynamics at zero 
temperature except that each spin keeps rotating in time 
and there is a preferred state. There are $q$ possible absorbing
states separated by phase even in this case. The approach
to absorbing state shows new exponents that are not seen
in known models.


\section{The model} 
We consider a $1D$ lattice of length $N$. The state at each site $i$ is 
denoted by $s_{t}(i)$. At $t=0$, we randomly assign each site a spin 
state $s_{0}(i)$ as $l$ where $l=1,2,\ldots k$ with equal probability.  
The states of the sites are updated synchronously in two steps.  

i) In the first step, we update the sites cyclically.  
We introduce one more
arbitrary variable $s'_{t}(i)$ such that $s'_{t+1}(i)={\rm{mod}}(s_{t}(i),k)+1$.

ii) In the second step, except for sites in a 
certain preferred state, all other sites mimic 
the state of a randomly chosen neighbour. 
In other words,  if $s'_{t+1}(i)\ne 1$, 
$s_{t+1}(i)=s'_{t+1}(i-1)$ or $s_{t+1}(i)=s'_{t+1}(i+1)$ 
with probability ${\frac{1}{2}}$.  
If $s'_{t+1}(i)=1$, we assign  $s_{t+1}(i)=1$. Thus, 
the site which is in state $1$ is not affected by 
its nearest neighbors. This rule breaks the symmetry between 
states. 
This rule models the possibility that some states are less affected 
by neighbours than others and all possible transitions are not equally probable.

For synchronization, we define an order parameter 
$O(t)={\frac{1}{N}}\sum_i H(s_{t}(i)-s_{t}(i-1))$ 
where $H(x)=0$ iff $x= 0$ and $H(x)=1$ otherwise. 
In other words, this order parameter is the number 
of domain walls where the spin values of consecutive 
sites on the lattice are different.  These can be
considered as fractions of unsatisfied bonds. A similar
quantity is known as energy density in Potts model with ferromagnetic coupling. 
The order parameter 
is zero for synchronized state and is nonzero otherwise.  
If the system goes to a synchronized state, the state changes in time but stays 
synchronized. Under stroboscopic observation, when we view the system only at times
that are multiples of $k$, the synchronized state is frozen. There are $k$ 
such possible states. Thus the ground state has $k$ fold degeneracy like
$q$ possible ground states in the $q-$Potts model. As mentioned above, there are models 
where infinite possible ground states, prerequisites of the Janssen-Grassberger conjecture
are not satisfied and the system is still in 
DP universality class. The ground state for the model introduced in this work has finite degeneracy.
The prerequisites of the Janssen-Grassberger conjecture are violated. We investigate if the DP class is robust for this case. It turns out to be a relevant perturbation and a new universality 
class is observed.

First, let us consider the case $k=3$.
In Fig. \ref{fig1}, we plot $O(t)$ as a function of time $t$ 
on a logarithmic scale. We find a power-law behaviour 
$O(t)\sim t^{-\delta}$ with $\delta\sim 0.778$ (All the fittings in work are computed using the fitting function of Gnuplot that implements nonlinear least-squares (NLLS) Marquardt-Levenberg algorithm. The fits are within 1$\%$.) There are fluctuations
of period $3$ over and above the power-law. 
In Fig. \ref{fig2}, we show how defects, (the sites where spins 
of consecutive sites are different) propagate in time.  
The defects propagate almost in a straight line and annihilate 
after meeting.
This behaviour is unlike the directed percolation universality class where defects propagate in a cone (See Fig.1 of \cite{janaki2003evidence}). We find that the defects take a 
preferred direction and there is very little random or zig-zag motion 
over and above this direction. 

We also study the evolution 
of the order parameter for different system sizes.  
In Fig. \ref{fig3} we show the evolution of the order 
parameter for different system sizes. There is a 
power-law behaviour followed by a cusp where the power changes. It is followed by decay. 
The cusp is observed at a time proportional to system size because the
defects move ballistically. if we plot $t/N^z$ 
versus $O(t)N^{z\delta}$ we observe an excellent 
collapse of the cusp which occurs around $t=6N$. Above the cusp, the power changes 
and is followed by decay. If we plot $t/N^{z'}$ 
versus $O(t)N^{z'\delta'}$ with $z'=2$ and $\delta'=0.5$, we observe an excellent 
collapse for $t>6N$. Thus there is a drift till $t \sim 6N$ and diffusion afterwards (See Fig. \ref{fig3}.)

So far, we have considered the case where all sites are randomly assigned to one of the 3 states. The question is what happens if the assignment is deterministic or random with non-uniform probability for different states. If 
$1\le i \le N/2$ are in one state and the rest of the lattice is in another state, there are 2 defects. These defects move almost ballistically, collide and merge. However, if the first $1/3$rd, second $1/3$rd and last $1/3$rd are in 3 different states, 3 defects move in the same direction. Two of them collide due to different effective speeds resulting from inherent stochasticity in the model. When there are two defects, they move in opposite directions, collide and merge.

\begin{figure}[h!]
 \centering
  \includegraphics[width=0.45\textwidth]{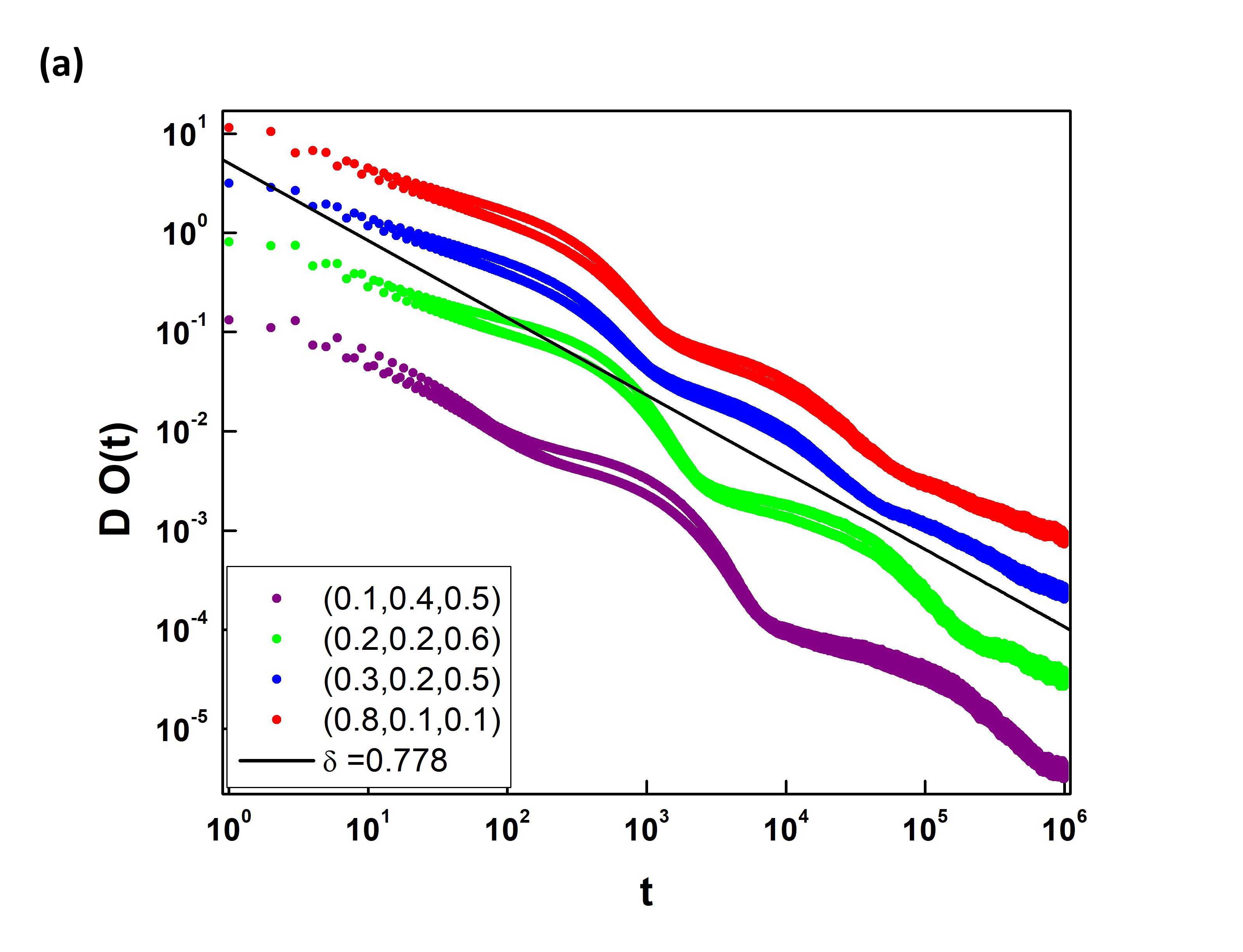}
  \includegraphics[width=0.45\textwidth]{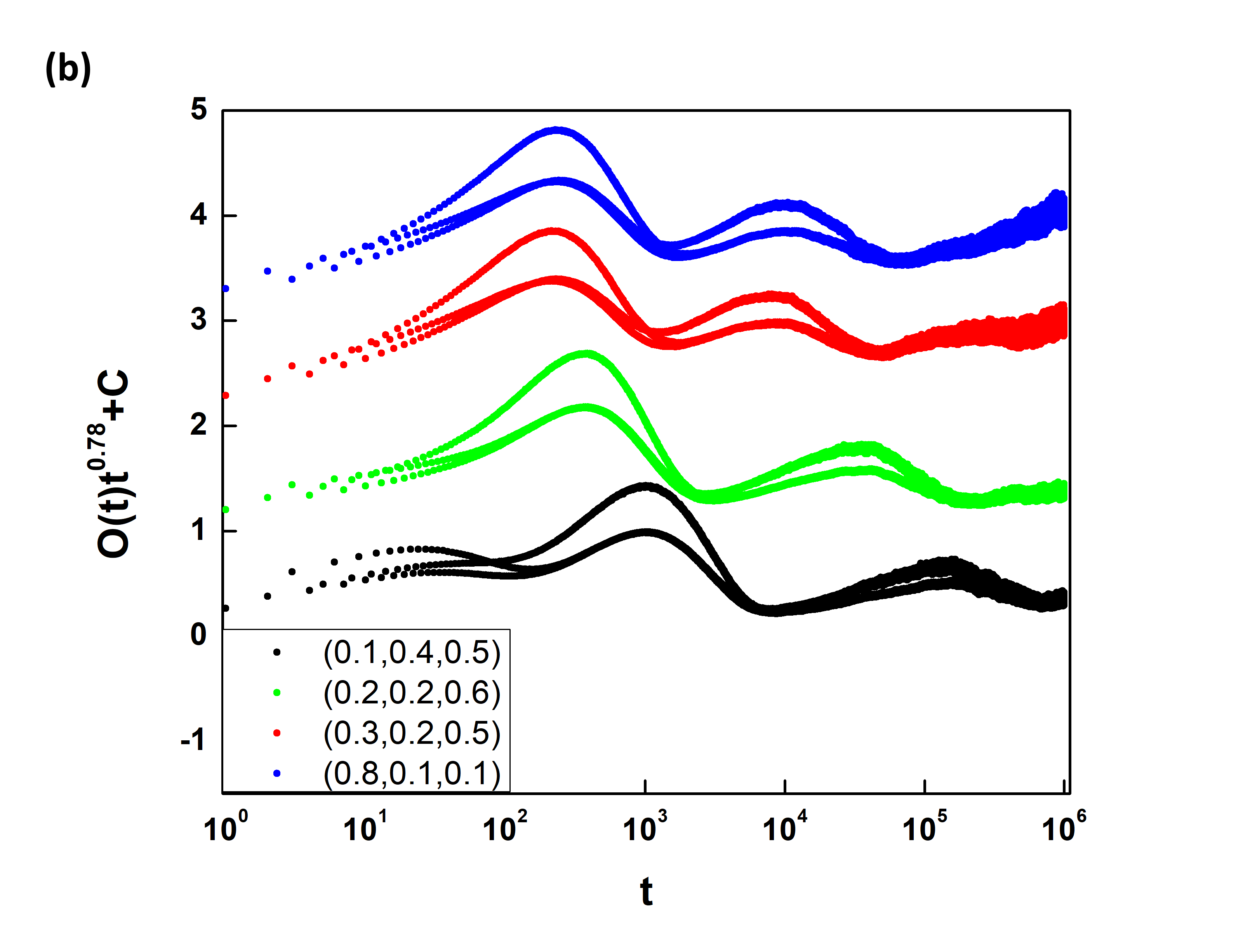}

 \caption{a) $O(t)$ is plotted as a function of $t$ for  $k=3$ for different initial
 probabilities denoted by 3-tuple $(P(1),P(2),P(3))$. The values on the y-axis are multiplied by arbitrary constants for
better visibility. We also plot the line proportional to $t^{-0.78}$ for reference. We 
demonstrate $(0.1,0.3,0.5)$, $(0.2,0.2,0.6)$, $(0.3,0.2,0.5)$ and $(0.8,0.1,0.1)$.
b) We plot $O(t) t^{\delta}+c$ for the above curves The oscillations dampen for long times
for the same 3-tuples. The curves are
appropriately shifted by arbitrary constant $c$ along the $y$ axis for better visibility. }
\label{fig8}
 \end{figure}

Now we consider the case of random but non-uniform 
distribution of three states. Let us define the probability of state $m$ at time $t=0$ by $P(m)$ where $1\le m\le 3$. The probability distribution of at $t=0$ can be denoted by a 3-tuple $(P(1),P(2),P(3))$. We consider the case  where none of the elements is 0. (If one of the states is absent, there is an exponential decay. If two states are absent, we start with a synchronized state.) In this case, we observe oscillations over and above the power-law. These oscillations could be log-periodic. However, it is difficult to ascertain it. The amplitude of 
oscillations decays in time. Thus, these oscillations may vanish asymptotically and there is a 
power-law decay with the same power as observed when $P(1)=P(2)=P(3)$. Some of the representative 
figures are shown in Fig. \ref{fig8}.

Now we study larger values of $k$ to check if the universality class changes for larger values of $k$. We studied $k=20, 50, 100$
and $k=200$. We carry out simulations $N=10^6$ and simulate for $3\times 10^6$. We observe a power-law decay over a range. The exponent changes only very marginally. It is possible that the exponent changes for a longer time. However, this study is beyond our computational limitations. The current simulations do not point to any significant change in $\delta$ with $k$.
\begin{figure}[h!]
 \centering
  \includegraphics[width=0.6\textwidth]{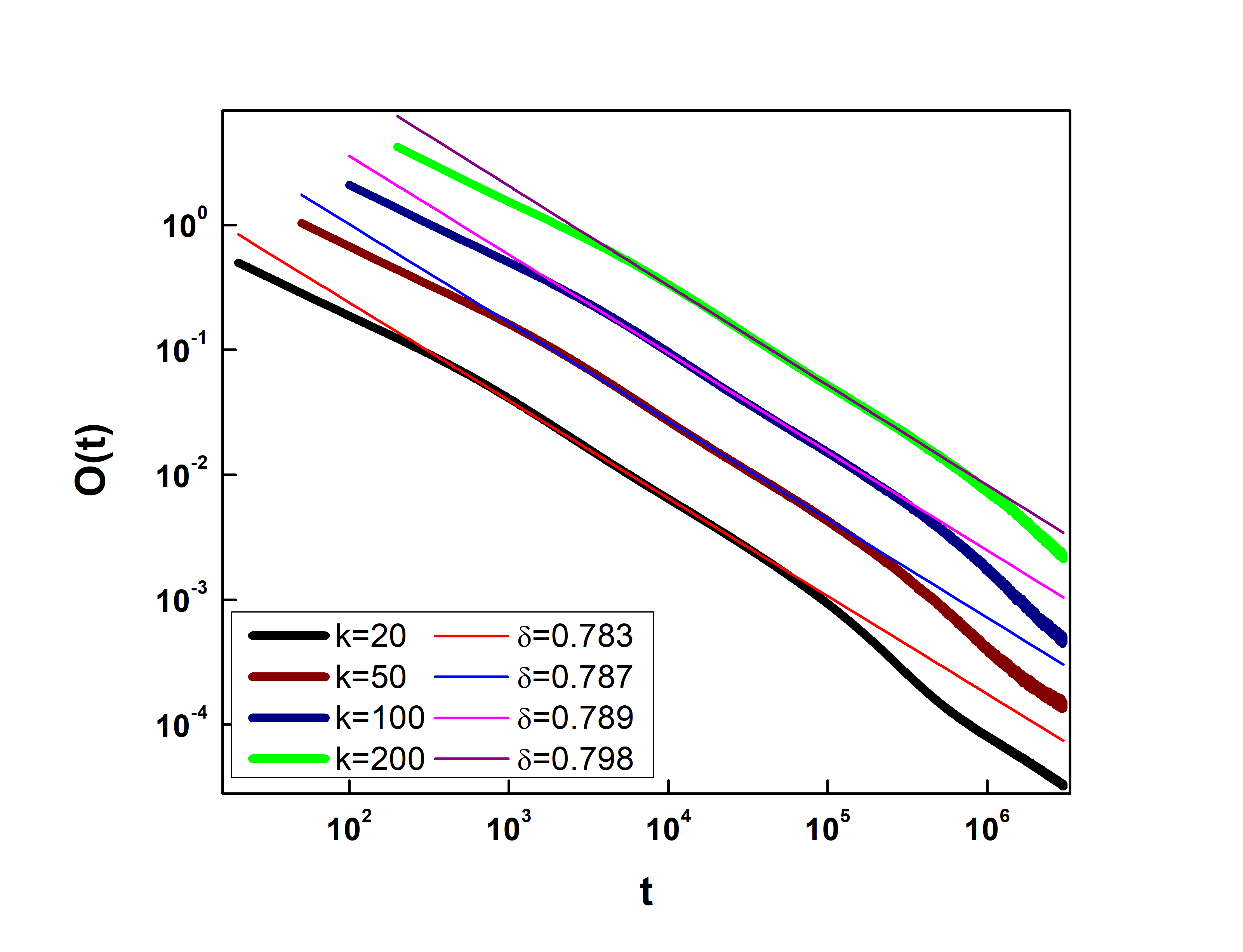}
 \caption{We plot $O(t)$ as a function of $t$  for  $k=20,50,100$ and 200 for $N=10^6$. The sites are assigned spin values with equal probability at $t=0$. We plot $O(t)$ only when $t$ is the exact multiple of $k$ because there are fluctuations of period $k$ over and above the power-law. The outputs are averaged for more than 1200 configurations. 
  The fits 
 proportional to $1/t^{\delta}$ with $\delta=0.783$, $\delta=0.787$, $\delta=0.789$ and $\delta=0.793$ are also shown. These fits over the relevant range are computed by the nonlinear fitting function of Gnuplot which implements the nonlinear least-squares (NLLS) Marquardt-Levenberg algorithm. The quantities on the y-axis are multiplied by arbitrary constant $D$ for better visibility.}
 \end{figure}

For $k=1$ there is no possibility of defects. Now we consider case $k=2$. For $k=2$, $O(t)\sim t^{-\delta}$ with $\delta\sim {\frac{1}{2}}$. (Fig. \ref{fig5}) If we investigate the time evolution of defects, it looks more like annihilating random walkers which explains the exponent ${\frac{1}{2}}$. (Fig. \ref{fig6}) We also study the size dependence of absorption for $k=2$ and we obtain excellent collapse if we plot $t/N^z$ versus $O(t)N^{z\delta}$ with $z=2$ and $\delta=1/2$ (Fig. \ref{fig7}).


\section{Discussion and Conclusion}
We have studied a model of cellular automaton for period-$n$ synchronization. We used cellular automaton to model a transition to the synchronized period-$n$ state. This model is updated in two steps. In the first step, the states of sites are updated cyclically. In the next step, the sites randomly take the state of their neighbour, except for sites with state 1. Sites with this state remain unaltered. We define the order parameter as the difference between the state values of the neighbouring states. The order parameter is a nonzero quantity in an unsynchronized state. In this work, we have studied various values of $k$ in detail. For $k=2$, it is observed that the defects grow as a power-law with the dynamic exponents $\delta=\frac{1}{2}$ and $z=2$. This suggests the model behaves like a random walk model for $k=2$. On the other hand, we get $\delta=0.778$  for $k=3$. For $k\ge3$, the model shows the same behaviour. The simulations do not point to any
significant change in the value of $\delta$ even for large $k$. These values are close to those recently observed in a transition to period-3 in the coupled Gaussian maps \cite{gaiki2024transition}. Finite-size studies show a cusp followed by decay. The time scales for cusp and decay scale differently with system size.  These exponents are novel and have not been seen in any other universality classes. Thus, these findings may suggest the existence of a new universality class.

\section{Acknowledgment}
PMG and DDJ thank DST-SERB for financial assistance (Ref. CRG/2020/003993). We thank Prof. D. Dhar and Prof. S. S. Manna for their suggestions. We thank Mr. Ankosh Deshmukh for his help with computation.




%



\section{References}
\bibliographystyle{elsarticle-num} 
\bibliography{ncycle.bib}

\end{document}